# Space very long baseline interferometry in China

Tao An, Xiaoyu Hong, Weimin Zheng, Shuhua Ye, Zhihan Qian,
Li Fu, Quan Guo, Sumit Jaiswal, Dali Kong, Baoqiang Lao, Lei Liu, Qinghui Liu,
Weijia Lv, Prashanth Mohan, Zhiqiang Shen, Guangli Wang, Fang Wu,
Xiaocong Wu, Juan Zhang, Zhongli Zhang, Zhenya Zheng, Weiye Zhong,
on behalf of the space VLBI team
Shanghai Astronomical Observatory, 80 Nandan Road, Shanghai 200030, China
antao@shao.ac.cn

June 4, 2019

**Abstract**

Space very long baseline interferometry (VLBI) has unique applications in high-resolution imaging of fine structure of astronomical objects and high-precision astrometry due to the key long space-Earth or space-space baselines beyond the Earth's diameter. China has been actively involved in the development of space VLBI in recent years. This review briefly summarizes China's research progress in space VLBI and the future development plan.

## 1   Background

Very long baseline interferometry (VLBI) is a modern observational technique based on combining geographically distributed radio telescopes to form an observation network, thereby enabling the highest possible angular resolution. The imaging resolution provided by VLBI is inversely proportional to the maximum baseline (the separation between two telescopes in the network) length and proportional to the observation wavelength. VLBI allows astronomers to observe the fine structure of compact celestial objects and acquire high-precision astrometry with sub-mas resolutions. Routinely operational VLBI networks, such as the European VLBI Network (EVN), the Very Long Baseline Array (VLBA) in the USA and East Asia VLBI Network (EAVN), have maximum baselines of ∼5000–10,000 km, and they yield high resolutions of 0.08–0.5 mas [An et al.(2018)], which are 100–600 times higher than those of the *Hubble* space telescope. However this is still not sufficient for some extremely compact unresolved objects. For example, imaging the black hole shadow of Sgr A*, the supermassive black hole (SMBH) at the heart of the Milky Way, requires a resolution of about 0.02 mas, which is on a scale comparable to the event horizon of the SMBH [Goddi, et al.(2017)].

Two conventional methods can be used to increase the imaging resolution: observing at highest possible frequency, and increasing the VLBI baseline length. The former method motivated the development of the Event Horizon Telescope [Doeleman(2017)] for directly observing the immediate environment of Sgr A* and M87 SMBHs at 230 GHz. The baseline lengths of ground-based VLBI networks are constrained by the size of the Earth. In order to acquire larger baseline length, astronomers have proposed sending radio telescopes into space to create a space-Earth VLBI network, *i.e.*, space VLBI.

Despite various technical difficulties and the huge associated costs, space VLBI can provide the highest resolution, as well as opening new observation windows in the electromagnetic spectrum by exploiting the lack of atmospheric disturbance in the space environment. Sub-millimeter (submm) wavelength and infrared observations from the Earth are highly constrained by the broad-band absorption of molecules in the atmosphere. Moreover, the turbulence in the atmosphere results in rapid fluctuations of the visibility phase to significantly reduce the coherence time for mm and submm wavelength interferometers, thereby decreasing the fringe detection sensitivity of VLBI. These factors motivated the development of space VLBI at short wavelengths for imaging the central SMBH shadow and studying planet formation. In addition, observations made from the ground below 10 MHz are strongly affected by absorption from the ionosphere, and thus this low frequency band is the last unexplored spectrum window. Even below 30 MHz, the turbulent ionosphere and the terrestrial radio frequency interference (RFI) signals severely affect the observational data received by ground telescopes. These low frequency observations are crucial for understanding the origin of the Universe, so the only opinion for making novel scientific discoveries is to deploy low-frequency radio telescopes or interferometers in space.

## 2   Space centimeter (cm) and mm wavelength VLBI arrays

Space VLBI was first proposed in 1970s (see recent reviews in : [Gurvits(2018), An et al.(2019)]). Two space VLBI missions for astronomy applications mainly at centimeter (cm) wavelengths comprised the VLBI Space Observatory Programme (VSOP) led by Japan [Hirabayashi et al.(1998)] and RadioAstron led by Russia [Kardashev et al.(2013)]. VSOP operated an 8-meter space telescope onboard the HALCA satellite, which was launched in 1997, and it worked for six years. The telescope was in an elliptical orbit with an apogee height of 21,400 km, thereby giving a maximum baseline about four times that of the VLBA to allow science investigations of active galactic nuclei (AGNs) and hydroxyl masers. Moreover, key technologies associated with space VLBI were tested. Fourteen years later, the RadioAstron with a 10-m telescope was launched in 2011 into a large elliptical orbit with an apogee height of 350,000 km. RadioAstron created a milestone with a record highest resolution of 8 $\mu$as [Baan et al.(2018), Kobalev et al.(2019)], which allowed the resolution of the emission zone in an AGN where the jet was launched, collimated and accelerated (e.g., [Giovannini, et al.(2018)]).

Following the success of VSOP and RadioAstron, Chinese astronomers have made rapid strides toward implementing the cm- and mm-wavelength space VLBI missions. Shanghai Astronomical Observatory belonging to the Chinese Academy of Sciences (CAS) undertook a concept study for space VLBI in early 2000s. The concept gradually evolved into a mission proposal for the space mm-wavelength VLBI array (SMVA: [Hong et al.(2014)]). A similar mission concept for next generation two-spacecraft space VLBI wass first introduced by NASA's Structure and Evolution of the Universe 2003 roadmap team (iARISE: [Murphy et al.(2005)]). In the SMVA roadmap, the proposed mission was planned with three stages, each with a differing science focus based on the expected implementation, as follow. (1) Long mm-wavelength space VLBI including two 10 m space telescopes working at frequencies up to 43 GHz. These two telescopes will obtain observations together with ground telescopes to achieve a highest resolution of 20 micro-arcsecond ($\mu$as). The main motivation is to obtain a better understand of AGN jet physics and the emission structure around SMBHs. (2) Short mm-wavelength (86 GHz) space VLBI involving three 12 m space telescopes. These telescopes could be operated either as a stand-alone space-only VLBI network or as a space-ground VLBI network. The atmosphere-free environment in space allows for a long integration time, but the shortcoming is the limited sensitivity of the 12 m space telescope compared with the large ground telescopes and the poor (u, v) coverage of only three elements. The telescopes can also be coordinated with ground-based mm-wanvelength telescopes to enhance the image quality by improving the (u, v) coverage. A key scientific aim is to explore the gas dynamics in the close vicinity of the event horizons of SMBHs. (3) Sub-mm space VLBI array comprising three to four 12-m telescopes. The implementation of multiple 12-m telescopes in space is expected to provide micro-Jy sensitivity with very high angular resolution at frequencies up to 230 GHz. These observational constraints are probably sufficient to detect proto-stellar disks during the formation of exoplanets.

Prototype studies related to the first stage of the SMVA have been faciliated by CAS grants for supporting scientific pre-research and investigating technical feasibility of the mission [Hong et al.(2014), Liu et al.(2016)]. A main achievement is the completion of a 10-m space antenna prototype illustrated in Fig. 1. Laboratory tests showed that the surface accuracy of this antenna is about 0.37 mm, thereby satisfying the 43-GHz observation requirement. Additional technical challenges related to the payloads need to be solved before the SMVA moves on to the engineering phase, including the cryogenic system of the receivers, highly stable time/frequency standard, high-rate data transfer to ground stations, and high-accuracy pointing performance of the space antenna. Chinese astronomers are still seeking opportunities to process these missions.

## 3   Space decameter to decimeter VLBI observatory

In addition to space VLBI strategies directed at producing high frequency observatories, such as the SMVA [Hong et al.(2014)] and Millimetron [Kardashev(2017)], another unique direction of space VLBI involves focusing on low frequencies (long wavelengths) but with higher sensitivity using large diameter space antennae. Many unique scientific studies are based exclusively on long-wavelength space VLBI. The long (decameter to decimeter) wavelength space VLBI experiences less technical difficulties regarding payloads compared with the submm wavelength space VLBI, and it can gain more support from the large (100-m level) telescopes on the Earth. Long wavelength observations are severely affected by absorption and refraction in the ionosphere, and high manmade RFI in ground-based telescopes (e.g., [Davies(1997)]), whereas these limiting factors are eliminated in space.

A remarkable advance was demonstrated in the latest Chinese Chang'E-4 mission based on three ultra-long-wavelength (ULW, <30 MHz) instruments [Jia et al.(2018)]. The first instrument is a micro-satellite (called Longjiang-2 and led by Harbin Institute of Technology, China) installed with a ULW dipole antenna, launched in 2018 May. This satellite, planned to orbit the Moon, performs radio observations by utilizing the RFI-clean environment at the far side of the Moon. The second ULW telescope is the collaborative Netherlands-China Low-Frequency Explorer, which is deployed on the relay satellite (called *Queqiao*) of Chang'E-4. This telescope will perform the first ULW space-ground VLBI experiments together with the ground-based LOw Frequency ARray (LOFAR: [van Haarlem et al.(2013)]). The third telescope led by Chinese astronomers is on the Chang'E-4 lander, which successfully landed on the far side of the Moon on January 2, 2019. These ULW telescopes are all in good condition at present and data are being collected. We look forward to their scientific findings. These ULW telescopes will start a new era for radio astronomy based on radio (including VLBI) observations at and below 30 MHz on and around the Moon. These pioneering missions provide the foundations for the implementation of future s scientific projects.

China has provided ongoing investment to support advances in space science in recent years [Wu & Bonnet(2017)]. Radio astronomy has also benefited from these rapid advances with exciting opportunities in space radio astronomy, and especially in space VLBI. Recently Chinese astronomers submitted a proposal for a low-frequency space VLBI programme called Cosmic Microscope (CM: [An et al.(2019)]) to the Chinese Academy of Sciences and the government of Shanghai. The planned CM program involves launching dual space telescopes with diameters of $\geq$ 30 m into large elliptical orbits with apogee heights of 90,000 and 60,000 km (see the sketch map in Fig. 2). A notable advantage of CM is the observational synergy with giant ground-based telescopes, e.g., Square Kilometre Array phase 1 (SKA1), Five hundred meter Aperture Spherical Telescope in China, and Arecibo 305 m, thereby allowing imaging of the ultra-low frequency radio sky at unprecedented high resolutions (0.4 mas at 1.67 GHz and 20 mas at 30 MHz) and with high sensivity at sub-mJy level. CM supports flexible observation modes (space-ground VLBI, space-space VLBI, and single space telescope) at four frequency bands ranging from 30 to 1670 MHz (18 cm to 10 m wavelengths). These capabilities allow CM to satisfy the needs of many scientific applications, including studies of the early Universe and cosmological structure formation, directly resolving SMBH binaries, precise positioning and distance measurement of pulsars and other transients, observation of hydroxyl masers in distant galaxies, and detection of magnetic fields from exoplanets. In the single-dish mode, each telescope can be used to monitor transient bursts and trigger follow-up VLBI observations. At large distances from the Earth, the telescopes may observe the neutral hydrogen lines originating from the epoch of cosmic dawn because the RFIs from the Earth become weaker with distance, thereby helping to measure the total angular power spectrum from the Epoch of Reionization and placing stringent constraints on cosmological models.

Based on the luminosity function of radio sources, the number of observable radio sources decreases sharply as the observational frequency increases, but the source count of extragalactic objects increase substantially in these low-frequency radio bands. Furthermore, steep-spectrum radio sources become brighter in the lower frequency bands ($\lesssim$ GHz). Thus the low-frequency space VLBI is expected to contribute greatly to statistical studies of the compact extragalactic radio source populations. Due to the increased sensitivity at and below 1 GHz, the future low-frequency space VLBI will play an essential role in understanding exotic celestial sources and transients, e.g., fast radio bursts [CHIME/FRB Collaboration et al.(2019)].

## 4   Space VLBI in the future

Previous space VLBI systems and related experimental satellites have successful obtained high-resolution astronomical observations to yield novel results and faciiate advances in observational, data processing and operational experience. The strategy for future development can be clearly stated as follows:

- First, to launch multiple space telescopes to improve the space-ground (u, v) coverage and unique space-space baselines. Previous space VLBI missions include a single space telescope, which resulted in a lack of sufficient space-ground baselines and no space-space baseline. The effects were apparent in the case of RadioAstron. During most of the satellite orbit, the large separation of <28 times Earth's diameter between RadioAstron and ground telescopes made the space-ground baselines produce a long narrow distribution in the (u, v) plane. This configuration resulted in strong sidelobes in the dirty image, which seriously affected the dynamic range of the image and it could also induce artifacts. Thus, new space VLBI programs, including the SMVA and CM, will generally employ two or more space telescopes to improve the (u, v) coverage, thereby enhancing the image quality. The difference precession rates for the orbital planes of various spacecraft will help to further improve the (u, v) coverage.

- The second aim is to increase the space-ground baseline sensitivity. The sensitivity of the space-ground VLBI baselines limited the previous space VLBI targets to only strong radio sources. This problem was related to the attitude adjustment systems of their satellites, which do not support rapid switching between sources. This has remained a key technical bottleneck and it has prevented space telescopes from conducting phase referencing observation, unlike ground telescopes. In addition, the diameter of space telescopes is 8–10 m. The best baseline sensitivity between the 10-m space telescope and the largest 100-m ground telescope is equivalent to two 30-m telescopes. A practical approach for detecting weaker sources is to enlarge the size of space telescopes by three or more times. For example, the baseline sensitivity between the proposed 30-m CM space telescope and the Effelsberg 100-m telescope will be three times higher than that between the existing 10-m space telescope and the Effelsberg 100-m telescope. Furthermore, considering that the SKA1 will be operational when the CM mission is funded and executed, the baseline sensitivity between the phased-up SKA1 and CM should be at least three times higher than that for the RadioAstron mission.

- The third goal is to develop space-only interferometers (including VLBI applications). As mentioned earlier, space-based VLBI is especially critical for the two spectrum windows that are not observable on the Earth comprising submillimeter-infrared and ultra low frequency bands. In order to full utilize space observations obtained with little or no influence from the Earth's atmosphere, an independently operated VLBI array comprising a number of space telescopes will be a specific model for space VLBI with unique scientific benefits. If these missions can be implemented, they will greatly advance key scientific goals, including SMBH imaging, transient studies, early Universe and exoplanet studies. Clearly, various technical bottlenecks related to space telescope design prevent the antenna size from being very large. The use of in-orbit splicing technology (e.g., [Buyakas et al.(1979), Gurvits(2000)]) will allow the production of a low-frequency telescope measuring 100 m or even 500 m. In the splicing technique, segments of a single large antenna are built and they are then joined together in space to produce a telescope using a robolic method. Several deployable telescopes or stations can be combined into a mini SKA in space.

- The final aim is to expand to Moon-based space VLBI. Many conventional studies of space VLBI included satellites which are either in the Earth's orbit or a large elliptical orbit around the Earth. However, advances in lunar exploration during the past decade mean that a Moon-based radio astronomical observatory is also under consideration (e.g., [Boonstra et al.(2016)]). In the latest Chang'E-4 mission, the lander is equipped with a low-frequency radio telescope and it landed on the far side of the Moon. Thus, radio environment data were measured on the far side of the Moon for the first time, which is an important milestone in Moon-based radio astronomy. The fourth phase of the China's lunar exploration program was initiated early in 2019. In around 2023, a relay satellite is planned for launch, which will be have an antenna with a diameter of about 4 m. This antenna will be responsible for data downlinking and it can also serve as a Moon-orbit VLBI station by working together with the VLBI network on the Earth. The primary objective of this VLBI system is to test and verify the key techniques associated with Moon-based VLBI, including payload design and data processing. Manned landing on the Moon provides another opportunity to deploy astronomical telescopes on the near side of the Moon. In addition, the far side of the Moon natually shields undesiable RFI signals from the Earth, so it is suitable for making low-frequency radio observations. Due to advances of space antenna technology, the large-scale deployment of lightweight low-frequency antennas on the far side of the Moon is expected to be feasible in the near future and they could eventually form a lunar-based SKA. The exploration of the Universe will be pushed to new levels by exploiting the pristine observational environment on the Moon.

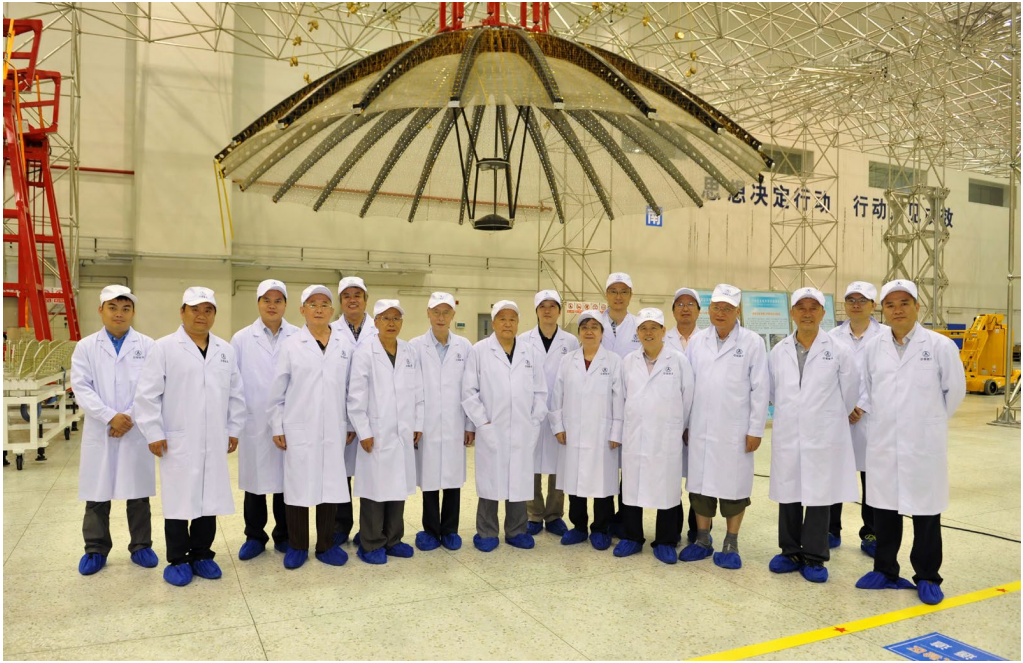

Figure 1: Image of the 10 m space antenna prototype of the mission (1) of the Space Millimeter-wavelength Very Array. It is complete in 2016. The surface accuracy is 0.37 mm.

## Acknowledgements

This work is funded by the National Key R&D Programme of China (2018YFA0404603) and Chinese Academy of Sciences (114231KYSB20170003). WMZ is grateful for funding support from the Ten Thousand Plan and the Program of Shanghai Subject Chief Scientist (14XD1404300). The author gratefully acknowledges the dedication of Chinese Space VLBI team, including hundreds of scientists, engineers, managers, and support personnel who continuously push forward the mission. TA thanks Leonid Gurvits for his careful proofreading and helpful comments on the manuscript.

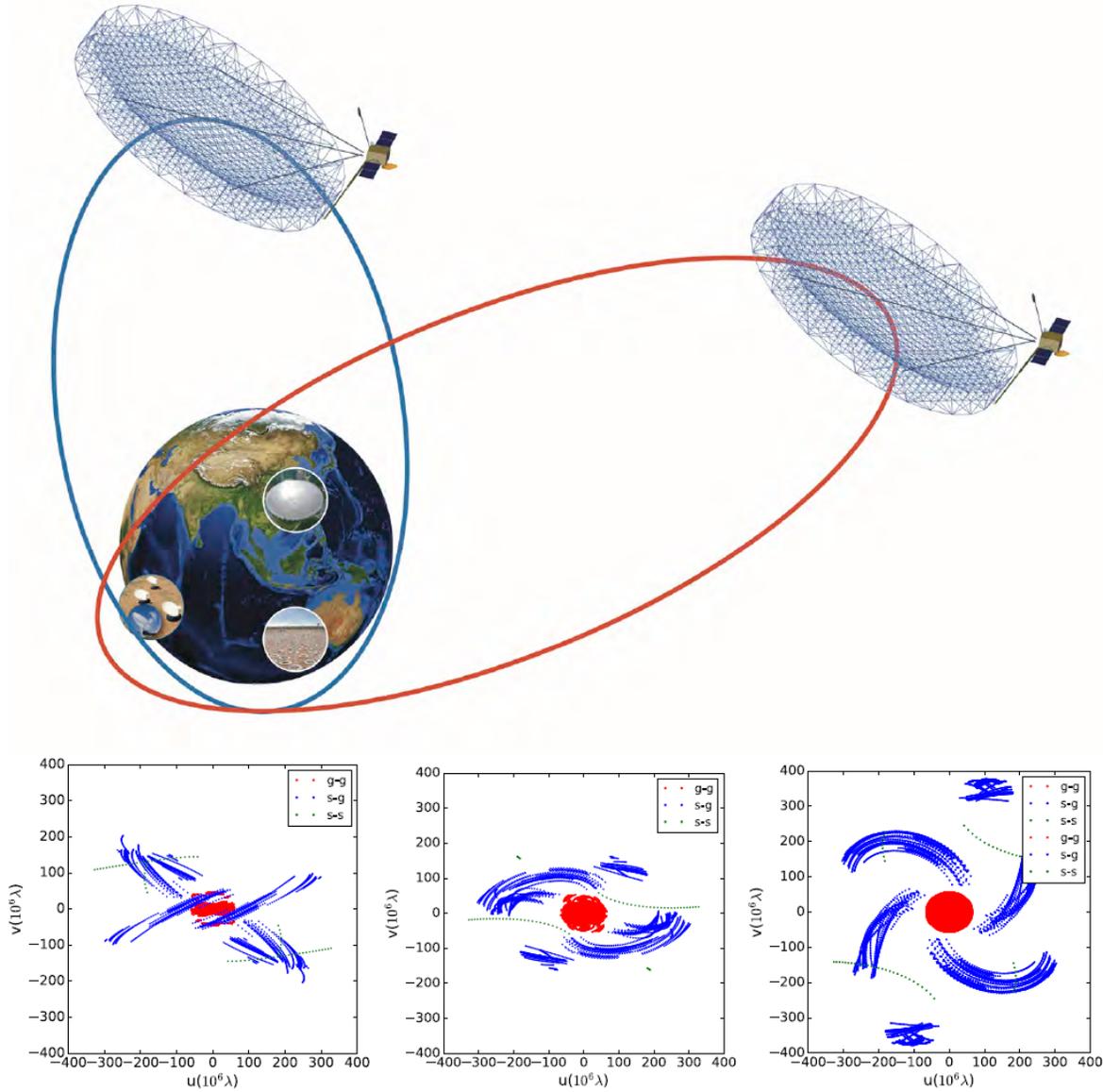

Figure 2: Sketch of the space-based dual telescopes of the Cosmic Microscope mission concept. It has been re-constructed from Figure 1 of [An et al.(2019)]. The two radio telescopes with diameter >30 m are launched into 2,000 km × 90,000 km, and 2,000 km × 60,000 km elliptical orbits. Compared to a single spacecraft, the dual spacecraft configuration improves the space-ground (u, v) coverage reducing sidelobes and increasing image quality. Copyright of the SKA telescope images: SKA Organisation. The bottom panel displays (*u,v*) coverages of three fake sources at declination of 0°, +30° and +60°, respectively. The observation frequency is 1.67 GHz. The participated ground telescopes are: European VLBI Network, Very Long Baseline Array of the USA and the Five hundred meter Aperture Spherical Telescope (FAST) of China. The orbit perigee is 2,000 km, and the apogee heights of two space telescopes are 60,000 km and 90,000 km. The initial inclination angle of the orbit is 28.5°. The longitude of the ascending node is 15°. Mention that these parameters are preliminary and they may change with the overall project design.
6